\documentclass{article}
\usepackage{LaThuileFPSpro}
\newcommand{\bi}{\bibitem}
\newcommand{\be}{\begin{eqnarray}}
\newcommand{\ee}{\end{eqnarray}}
\newcommand{\rar}{\rightarrow}
\begin{document}
\title{ 
COSMIC ANTIMATTER: MODELS  AND OBSERVATIONAL BOUNDS
  }
\author{
A.D. Dolgov \\
  {\em ITEP, 117218, Moscow, Russia}\\
{\em INFN, Ferrara 40100, Italy}\\
{\em University of Ferrara, Ferrara 40100, Italy}\\
   }
\maketitle

\baselineskip=11.6pt

\begin{abstract}

A model which leads to abundant antimatter objects in the Galaxy (anti-clouds, 
anti-stars, etc) is presented. Observational manifestations are analyzed. In 
particular, the model allows for all cosmological dark matter to be made out 
of compact baryonic and antibaryonic objects.

\end{abstract}
\newpage
\section{Introduction}
The origin of the observed excess of matter over antimatter in the universe
is believed to be pretty well understood now. As formulated by 
Sakharov~\cite{ads}:\\
1) nonconservation of baryonic number,\\
2) breaking of C and CP, and\\
3) deviation from thermal equilibrium \\
lead to different cosmological abundances of baryons and antibaryons.
The cosmological baryon asymmetry is characterized by the dimensionless
ratio of the difference between the number densities of baryons and antibaryons
to the number density of photons in the cosmic microwave background radiation:
\be
\beta =  \frac{n_B - n_{\bar B}}{n_\gamma} \approx 6\cdot 10^{-10}
\label{beta}
\ee
There are many theoretical scenarios which allow to ``explain'' this value
of the baryon asymmetry, for the review see~\cite{bs-rev}. Unfortunately
``many'' means that we do not know the single one (or several?) of the 
suggested mechanisms which was indeed realized. Usually in such cases
experiment is the judge which says what is right or wrong. However, it
is impossible to distinguish between 
competing mechanisms having in one's disposal
only one number, the same for all the scenarios. We would be in much better
situation if $\beta$ is not a constant over all the universe but is a 
function of space point, $\beta = \beta (x)$. So it is interesting
to study the mechanisms which might lead to space varying $\beta$ and 
especially, in some regions of space, to $\beta < 0$, i.e. to possible 
generation of cosmological antimatter. 

There is an increasing experimental activity in search for cosmic antimatter.
In addition to the already existing detectors, BESS, Pamella, and AMS, a few more
sensitive ones shall be launched in the nearest years, AMS-02 (2009),
PEBS (2010), and GAPS (2013), see the review talk~\cite{picozza} at TAUP 2007. 
To the present time no positive results indicating an astronomically significant 
cosmic antimatter have been found but still the bounds are rather loose and
as we see in what follows, it is not excluded that the amount of antimatter
in the universe may be comparable to that of matter and astronomically large
antimatter objects can be in our Galaxy quite close to us. 

If this is the case, one should search and
may hope to observe cosmic antinuclei starting 
from $ ^4He $ to much heavier ones, excessive antiprotons and positrons,
flux of energetic gamma rays with energies about 100 MeV from
$p \bar p$--annihilation and 0.511 MeV from $e^- e^+$--annihilation,
violent phenomena from antistars and anticlouds, and some other
more subtle ones.

We cannot say, of course, if there is any reasonable chance to find
all that, but at least there is a simple theoretical model according
to which galaxies, including the Galaxy, though possibly dominated 
by matter, may include astronomically significant clumps of antimatter
on the verge of possible detection.

This talk consists of the following two main parts: \\
I. The mechanism of the antimatter creation leading to considerable
amount of antimatter in the Galaxy in the form of compact objects or
clouds. \\
II. Antimatter phenomenology, observational signatures, and bounds.\\

The talk is based on several papers written in collaboration with
C. Bambi,  M. Kawasaki, N. Kevlishvili, and 
J. Silk~\cite{ad-js,cb-ad,ad-mk-nk}, where a detailed
discussion and more complete list of references can be found.

\section{Standard homogeneous baryogenesis and bounds on antimatter}

Up to now we have observed only matter and no antimatter, except for a
little antiprotons and positrons most probably of secondary origin.
However, the observed intensive 0.511 MeV line from the galactic 
center~\cite{ann-line}, which surely originated from the electron-positron 
annihilation, $e^+ e^- \rar 2\gamma$, may be a signature of cosmic antimatter. 
Still astronomical data rather disfavor cosmologically significant amount of
antimatter. In our neighborhood the nearest anti-galaxy may be at least
at the distance  of 10 Mpc~\cite{steigman}. This result can be obtained as
follows. At such distance the antigalaxy should be in the same cloud of
intergalactic gas as e.g. our Galaxy. 
The number of annihilation per second of the intergalactic gas inside such 
antigalaxy can be estimated as:
 \be
 \dot N = \sigma_{ann} v N_{gal} \langle n_{p} \rangle = 
 10^{47}/{ {sec}}  
\label{dot-N}
\ee
where ${ \sigma_{ann} v = 10^{-15}} $ cm${^3}$/s, 
$ {N_{gal}  \sim 10^{67} }$ is the total number of antiprotons in the gas 
which is contained in the antigalaxy, 
${  \langle n_{p} \rangle \sim 10^{-5} }$/cm${ ^3}$ is the number density of protons
in the intergalactic gas. The gamma ray luminosity produced by the
annihilation is $ L = 10^{43}\,{\rm erg/s}.$ It would create
the constant in time energy flux on the Earth, $ F = 10^{-3} $ MeV/cm${\bf ^2}$/s,
which is excluded by observations. For comparison, the typical (short-time) flux
from the gamma-bursters is about $ 10^{2}$ MeV/cm${ ^2}$/s. 

There are observed colliding galaxies at larger distances. They should consist
of the same kind of matter (or antimatter?). If galaxy and antigalaxy collide the
gamma-ray luminosity would be 5 orders of magnitude higher (proportional to the 
number density of gas inside galaxies) than the luminosity in the case of 
antigalaxy washed by the intergalactic gas. This allows to conclude that 
colliding galaxy and antigalaxy should be at 300 times larger distance, i.e.
at or outside the present day cosmological horizon.

Esthetically attractive is the charge symmetric cosmology, with equal weight
of cosmologically large domains of matter and antimatter. Such situation
is almost inevitable if CP is spontaneously broken~\cite{lee}. It was shown,
however, that in charge symmetric universe the nearest antimatter domain should 
be at the distance larger than a Gpc~\cite{cdg}, because the matter-antimatter  
annihilation at the domain boundaries would produce
too intensive gamma ray background. 

So we have to conclude that an asymmetric production of matter and
antimatter is necessary. In the model considered below it is almost 
symmetric but the bulk of baryonic and/or antibaryonic matter
can escape observations if antimatter ``lives'' in compact
high density objects. Observational restrictions on astronomically
large but subdominant antimatter objects/domains, anti-stars, anti-clouds,
etc, are rather loose and strongly depend upon the type of the objects.

\section{Anti-creation mechanism}

The model which leads to creation of an almost baryosymmetric universe
with the bulk of matter in the form of relatively compact objects consisting
of baryons and antibaryons was put forward in ref.~\cite{ad-js} and recently
further developed in~\cite{ad-mk-nk}. The model is based on the slightly 
modified version of the Affleck-Dine (AD) baryogenesis scenario~\cite{affl-dine}.
According to AD scenario a very large baryon asymmetry of the universe might
be generated due to accumulation of baryonic charge along flat
directions of the potential of a scalar field $\chi$ with nonzero baryonic number.
Normally very high $\beta \sim 1$ is predicted and theoretical efforts
are needed to diminish the result. However, if the window to the flat directions
is open only during a short period, cosmologically small but possibly
astronomically large bubbles with high $\beta$ could be created, while
the rest of the universe would have the normal $\beta \approx 6\cdot 10^{-10}$.
Such high $B$ bubbles would 
occupy a small fraction of the universe volume, but may make a dominant
contribution to the total mass of the baryonic matter. They can even make
all cosmological dark matter in the form of compact already dead (anti)stars
or primordial black holes (PBH).

To achieve this goal one should add a general renormalizable coupling of the scalar
baryon $\chi$ to the inflaton $\Phi$:
\be 
U_\chi (\chi, \Phi) = \lambda_1(\Phi-\Phi_1)^2|\chi|^2
+\lambda_2|\chi|^4\ln{\frac{|\chi|^2}{\sigma^2}}+ m_0^2 |\chi|^2
+ m_1^2\chi^2+m_1^{*2}\chi^{*2}.
\label{U-of-chi-Phi}
\ee
where $\Phi_1$ is some value of the inflaton field which it passes closer to the
end of inflation. Its value is chosen so that after passing $\Phi_1$ inflation is
still significant to make large B-bubbles. The second term in the potential is
Coleman-Weinberg potential~\cite{Coleman:1973} which is obtained by 
summation of one loop corrections to the quartic potential, $\lambda_2 |\chi|^4$.
The last two mass terms are not invariant with respect to the phase
rotation:
\be 
\chi \rightarrow e^{i\theta} \chi
\label{chi-transform}
\ee
and thus break baryonic current conservation. It can be seen from the
following mechanical analogy. The equation of motion of homogeneous field $\chi(t)$:
\be 
\ddot \chi + 3 H \dot \chi + 
\frac{\partial U(\chi,\Phi)}{\partial \chi^*} = 0
\label{ddot-chi}
\ee
is just the equation of motion of point-like particle in potential $U$ with
the liquid friction term proportional to the Hubble parameter $H$. In this 
language the baryonic number, which is the time component of the current
\be
{J_\mu^{(B)} = i\chi^\dagger \partial_\mu \chi + h.c. },
\label{j-mu}
\ee
is the angular momentum of this motion. If the potential is spherically
symmetric i.e. it depends upon $|\chi|$, angular momentum is conserved.
The last two terms break spherical symmetry and give rise to B-nonconservation.

Depending upon the value of $\Phi$, potential $U(\chi,\Phi)$ has 
either one minimum at $\chi =0$, or two minima: at $\chi= 0$ and
some $\chi_2(\Phi) \neq 0$, or again one minimum at $\chi_2(\Phi)$,
see fig. 1.

\begin{figure}[t]
  \vspace{9.0cm}
  \includegraphics{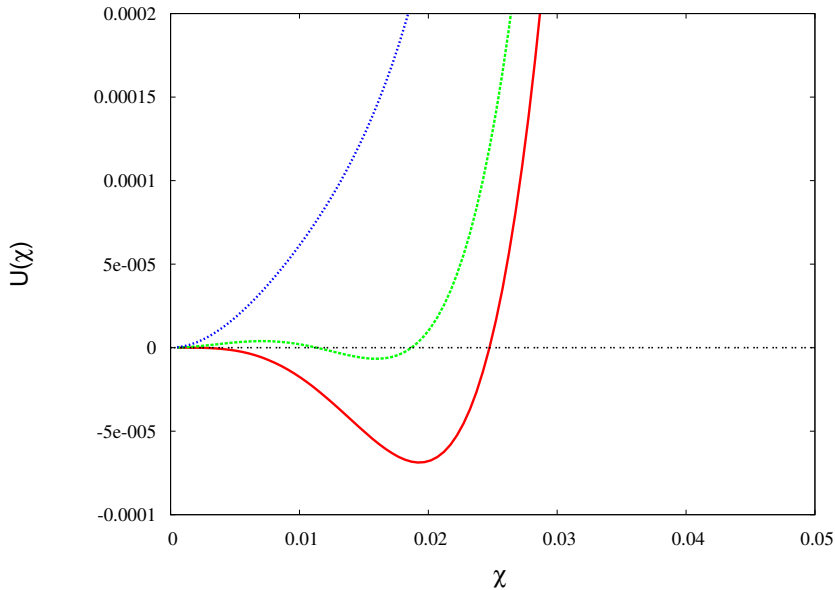}
  \caption{\it The evolution of  potential $U(\chi,\Phi)$ for different values
of the inflaton field $ \Phi$.
        \label{U-of-chi} }
\end{figure}

The behavior of $\chi$ in this potential is more or less evident. When the
potential well near the minimum at $\chi=0$ becomes low, the field can
quantum fluctuate away from zero and if $\chi$ reaches sufficiently large
magnitude during period when the second deeper minimum at $\chi_2$ exists,
it would live there till this second minimum disappears. Otherwise $\chi$
would remain at $\chi = 0$. Choosing the parameters of the potential
we can make the probability to fluctuate to the second minimum sufficiently 
small. When the minimum at $\chi_2$ disappears $\chi$ would move down to zero
oscillating around it with decreasing amplitude. The decrease is due to
the cosmological expansion and to particle production by the oscillating 
field $\chi$. The evolution of $\chi$ is presented in fig. 2, according
to numerical calculations of ref.~\cite{ad-mk-nk}.

\begin{figure}[t]
  \vspace{9.0cm}
  \includegraphics{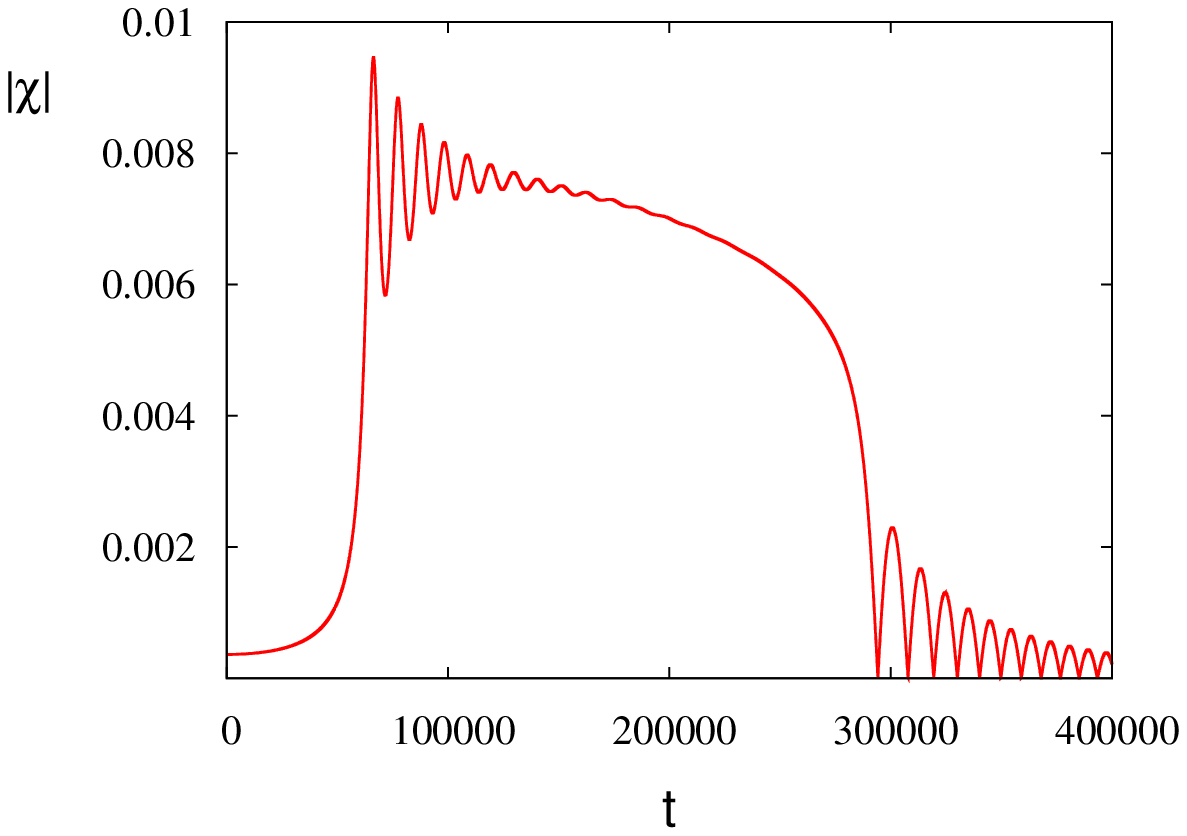}
  \caption{\it Evolution of $|\chi|$ because of the shift of
the position of the second minimum in $U(\chi,\Phi)$.
        \label{chi-evolv} }
\end{figure}

An important feature of the solution is the rotation of $\chi$ around the
point $\chi =0$, induced by the non-sphericity of the potential at low
$\chi$. As is argued above, this rotation is just non-zero baryonic 
charge density of $\chi$. Baryonic number stored in this rotation 
is transformed into excess of quarks over antiquarks or vice versa by 
B-conserving $\chi$ decays.

The magnitude of the baryon asymmetry, $\beta$, inside the bubbles which 
were filled with large $\chi$ (B-balls) and the bubble size are stochastic 
quantities. The initial phase, $\theta$, is uniform in the interval
$[0,2\pi]$ since due to the large Hubble term, $H\gg m_1$, quantum
fluctuations equally populate the circle of the second minimum of
$U(\chi,\Phi)$ (\ref{U-of-chi-Phi}) where $\chi = \chi_2$. The generated
baryonic number (angular momentum) is proportional to the displacement
of the phase with respect to the valley where 
$m_1^2 \chi^2 + m_1^{*2} \chi^{*2}$ has minimal value. Evidently the bubbles
with negative and positive $\beta$ are equally probable. The magnitude of
the asymmetry inside B-bubbles is also uniformly distributed in the interval
$[-\beta_m, \beta_m]$, where $\beta_m$ is the maximum of the asymmetry which
may be of the order of unity. The baryon asymmetry inside the bubbles
can be especially large if $\chi$ decayed much after the inflaton decay.
In this case the cosmological energy density would be dominated by 
non-relativistic $\chi$ prior to its decay
and all the baryonic number would be normalized
to photons produced by $\chi$ decay products only.

A simple modification of the potential $U(\chi,\Phi)$ (\ref{U-of-chi-Phi})
can shift the matter-antimatter symmetry of B-bubble population in either
way and magnitude, see e.g.~\cite{ad-b-asym}. In this way the universe
with the homogeneous background baryon asymmetry $\beta = 6\cdot 10^{-10}$ 
and small regions with $\beta \sim 1$ of both signs can be created.
Despite a small fraction of the volume, B-bubbles may dominate in the
cosmological energy density.

The size of B-ball is determined by the remaining inflationary time
after inflaton field passed $\Phi_1$ and can be as large as the solar mass or
even much larger, or as small as $10^{15}-10^{20}$ g or even smaller.

According to the calculations of refs.~\cite{ad-js,ad-mk-nk} the initial 
mass spectrum has a very simple log-normal form:
\be 
\frac{dN}{dM} = C\,exp \left[ -\gamma\, \ln^2 \left(\frac{M}{M_1}\right)
\right],
\label{dN-dM}
\ee
where C, ${\gamma}$, and ${ M_1}$ are unknown constant parameters.
If ${M_1 \sim M_\odot}$ some of these high $\beta$ bubbles 
might form stellar type objects and primordial black holes (PBH). 
With much smaller $M_1$ light PBHs, but still
with sufficiently large masses to save them from the Hawking evaporation 
during the universe life-time, could be created. 
Relatively light PBH with $M \approx 10^{17}$ g and
mass spectrum (\ref{dN-dM}) may be the source of 0.511 line from
$e^+e^-$--annihilation~\cite{cb-ad-ap}, observed in the galactic center.
In all the cases of heavy or light PBH and/or evolved, now dead or low 
luminosity, stars, they could make (all) cosmological dark matter.  

Due to subsequent accretion of matter the initial spectrum 
(\ref{dN-dM}) would be somewhat distorted. The calculations are in
progress but here in phenomenological application we assume that the 
spectrum is not modified.

\section{Inhomogeneities}
In this scenario there two mechanisms of creation of density perturbations
at small scales:\\
1. After formation of domains with large ${\chi}$ the equation of
state inside and outside of the domains would be different. Inside the
domains $\langle \chi \rangle \neq 0$ and the equation of state approaches
the nonrelativistic one, while outside the domains the equation of state
remains relativistic for a long time. As is known, in this case 
isocurvature perturbations are generated which in the course 
of evolution are transformed into real density perturbations with
$\delta \rho \neq 0$.\\
2. After the QCD phase transition at ${ T\sim 100}$ MeV, when quarks made
non-relativistic protons, the matter inside B-balls would quickly become 
nonrelativistic and a large density contrast could be created.

As we just have mentioned the initially inhomogeneous $\chi$ and/or
$ \beta$ lead to isocurvature perturbations. The
amplitude of such perturbations is restricted by CMBR at about 10\%
level, but the bounds from CMBR are valid at quite large wave lengths,
larger than  $\sim 10$ Mpc.

If ${\delta\rho /\rho = 1}$ at horizon crossing, PBHs
could be formed. The mass inside the horizon as a function of the
cosmological time is:
\be
{{ M_{hor} = 10^{38} \rm{ g}\, ({ t}/\rm{ sec})}}
\label{M-hor}
\ee
For relativistic expansion regime time is related to temperature as
$t({\rm sec}) \approx 1/T^2({\rm MeV})$. Thus   
for ${T=10^{8}}$ GeV at the horizon crossing the PBH mass would be 
${10^{16} }$ g. At the QCD phase transition and below the mass inside
the horizon can be from the solar mass up to 
${10^{6-7} M_\odot}$ on the tail of the distribution. 
This presents a new mechanism of an early quasar formation which 
naturally explains their large masses already at high red-shifts and 
their evolved chemistry.

Anti-BH may be surrounded by anti-atmosphere if $ \beta$ {slowly} 
decreases. There is no observational difference between black holes
and anti black holes but the atmosphere may betray them

The masses may be even larger than millions solar masses, 
but we assume that $M_0$ in eq. (\ref{dN-dM}) does not
exceed a few solar masses, so the formation of BHs much more massive than
indicated above is strongly suppressed.  
Compact objects (not BH) with smaller masses might be formed too depending 
upon the relation between their mass and the Jeans mass (see below).

The density contrast created by an almost instant transformation of
relativistic quarks into nonrelativistic baryons is equal to:
\be 
r_B=\frac{\delta\rho}{\rho} = \frac{\beta n_\gamma m_p}{(\pi^2/30) g_* T^4}
\approx 0.07 \beta\,\frac{ m_p}{T}. 
\label{r-B}
\ee
The nonrelativistic baryonic matter started to dominate inside the bubble 
at the temperature: 
\be {{
T =T_{in} \approx 65 \, \beta \,{\rm MeV}
}}\label{T-of-beta}
\label{T-eq}
\ee
The mass inside a baryon-rich bubble at the radiation dominated stage is 
\be {{{
M_B \approx
2\cdot 10^5 \, M_\odot (1+r_B) \left(\frac{R_B}{2t}\right)^3\,
\left(\frac{t}{\rm sec}\right) }}}
\label{M-B}
\ee
The mass density in such a bubble at the onset of matter domination is
\be{{
\rho_B 
\approx 10^{13} \beta^4 \,\, {\rm g/cm}^3\, .
}}\label{rho-B}
\ee
When a B-bubble entered under horizon its evolution in the early universe 
is determined by the relation between its radius, $R_B$ and the Jeans wave 
length, $\lambda_J$. The latter at the onset of MD-dominance is
\be 
\lambda_J = c_s \left({\frac{\pi M_{Pl}^2}{\rho}}\right)^{1/2} 
\approx 10 t\,\left(\frac{T}{m_N}\right)^{1/2} 
\label{lambda-J}
\ee
where the speed of sound is taken as 
${{ c_s \approx \left({T/m_N}\right)^{1/2}}}$.

The bubbles with ${{ {\delta\rho}/{\rho}<1}}$ but with
$R_B> \lambda_J$ and correspondingly ${{M_B>M_{Jeans}}} $
at horizon would decouple from cosmological expansion and
form compact stellar type objects or ``low'' density clouds.
For further implication it is important to know
what anti-objects could survive against an early annihilation?

The initial value of the Jeans mass is equal to:
\be{{{
M_J \approx 135\left(\frac{T}{m_N}\right)^{3/2} M_{Pl}^2 t
\approx 100 \frac{M_\odot}{ \beta^{1/2}}
}}}\label{M-J}
\ee
Taken literally this expression leads to a slow, as ${1/\sqrt{T}}$,
increase of $ {M_J}$ and ${\lambda_j}$. However, this is not so
because in a matter dominated object with a high baryon-to-photon
ratio the temperature drops as ${T\sim 1/a^2}$ and $M_J$ decreases too:
${M_J \sim 1/a^{3/2}}$. For example, for B-balls with
approximately solar mass ${ M_B\sim M_\odot}$ and the radius
$ {R_B \approx 10^9}$ cm at horizon crossing the mass density 
behaves as:
\be {{
\rho_B = \rho_B^{(in)} (a_{in}/a)^{3} \approx 6\cdot 10^5\,\,
{\rm g/cm^3}}}.
\label{rho-Bb}
\ee
The temperature inside such a B-ball at the moment when ${M_J=M_\odot}$
is equal to: 
\be {{
T\approx T_{in} (a_{in}/a)^2 \approx 0.025\,\,{\rm  MeV}. 
}}\label{T-msun}
\ee
Such an object is similar to the red giant core.

\section{ Universe heating by B-balls }
There are three processes of energy release which are potentially 
important for B-ball survival and for the physics of the early 
universe (BBN, CMBR, reionization, etc):\\
{1. Cooling down of B-balls because of their
high internal temperature.\\
{2. Annihilation of the surrounding matter on the surface.}\\
{3. Nuclear reactions inside.} \\
We will briefly discuss them in what follows.
\\

1. Initially the temperature inside B-balls was smaller
than the outside temperature because of faster cooling of
nonrelativistic matter. So such stellar-like object were formed
in the background plasma with higher temperature and higher 
external pressure. It is in a drastic contrast with normal stars
where the situation is the opposite.

After the B-bubble mass became larger
than the Jeans mass, the ball expansion stopped and the internal
temperature gradually became larger than the external one and 
B-balls started to radiate into external space.
The cooling time is determined by the photon diffusion:
\be {{
t_{diff} 
\approx 2\cdot 10^{11}\,{\rm sec} 
\left(\frac{M_B}{M_\odot}\right)\,\left(\frac{\rm sec}{R_B}\right) 
\left(\frac{\sigma_{e\gamma}}{\sigma_{Th}}\right)
}}
\label{t-diff}
\ee
The thermal energy stored inside B-ball is
\be {{
E_{therm}^{(tot)} = 3T M_B / m_N 
\approx  1.5\cdot 10^{50} {\rm erg}
}}\label{E-therm}
\ee
and the luminosity determined by the diffusion time (\ref{t-diff})
would be {${L\approx 10^{39}}$ erg/sec.} 

If B-balls make all cosmological dark matter, their fraction cannot 
exceed ${ \Omega_{DM} = 0.25}$. Hence the thermal keV photons would make
{${(10^{-4}-10^{-5})\Delta}$ of CMBR,} red-shifted today to the 
background light. Here $\Delta$ is the fraction of B-balls 
with solar mass and $\sim$keV internal temperature.

2. If B-ball is similar to the red giant core
the nuclear helium burning inside would proceed through the reaction
$ {3He^4 \rar C^{12}}$, however with larger T by the factor
$ {\sim 2.5}$. Since the luminosity with respect to this
process strongly depends upon the temperature,
$ {L\sim T^{40}}$, the life-time of such B-ball
would be very short. The total energy influx from such
B-ball would be below
${10^{-4}}$ of CMBR if {${\tau < 10^9}$ s.} The efficient
nuclear reactions inside B-balls could lead to B-ball explosion and 
creation of solar mass anti-cloud which might quickly disappear due
to matter-antimatter annihilation inside the whole volume of the cloud.
It is difficult to make a qualitative conclusion without detailed
calculations.

3. For compact objects, in contrast to clouds, the
annihilation could proceed only on the surface and they would have
much longer life-time. The (anti)proton mean free path before 
recombination is small:
\be{{{
l_p = \frac{1}{(\sigma n)} \sim \frac{m_p^2}{\alpha^2\,T^3} = 0.1\, cm\, 
\left(\frac{MeV}{T}\right)^3
}}\label{l-p}
}\ee
and the annihilation can be neglected. After recombination the number of 
annihilation on one B-ball per unit time would be: 
\be
\dot N = 
10^{31} V_p \left(\frac{T}{ 0.1\,\,{ {eV}}}\right)^3
\left(\frac{R_B}{10^9\,\,{{ cm}}}\right)^2,
\label{dot-N-recomb}
\ee
The energy release from this process would give about ${10^{-15}}$ of 
the CMBR energy density.

\section{Early summary}
1. Compact anti-objects mostly survived in the early universe.\\
{2. A kind of early dense stars might be formed with 
initial pressure outside larger than that inside.}\\
{3. Such ``stars'' may evolve quickly and, in
particular, make early SNs, enrich the universe with heavy 
(anti)nuclei and re-ionize the universe.} \\
{4. The energy release from stellar like objects in the early
universe is small compared to CMBR.}\\
5. B-balls are not dangerous for BBN since the volume of 
B-bubbles is small. Moreover, one can always hide any undesirable 
objects into black holes.\\[2mm]
For more rigorous conclusion detailed calculations are necessary.


\section{Antimatter in contemporary universe }

Here we will discuss phenomenological manifestations of possible
astronomical anti-objects which may be in the Galaxy. We will
use the theory discussed above which may lead to their creation
as a guiding line but will not heavily rely on any theory for 
the conclusions. We assume that anything which is not forbidden 
is allowed and consider observational consequences of such practically
unrestricted assumption. \\[1mm]
Astronomical objects which may live in our neighborhood include:\\
{1. Gas clouds of antimatter.}\\
{2. Isolated antistars.}\\
{3. Anti stellar clusters.}\\
4. Anti black holes.\\
{5. Anything else not included into the list above.}\\
Such objects may be: inside galaxies or outside galaxies,
inside galactic halos or in intergalactic space. We will
consider all the options.

\subsection{Photons from annihilation}

The observational signatures of these (anti)objects
would be a 100 MeV gamma background, excessive antiprotons and
positrons in cosmic rays, antinuclei,
compact sources of gamma radiation, and probably more difficult,
a measurement of photon polarization from synchrotron radiation and
fluxes of neutrino versus antineutrino in neutrino telescopes.

Astronomically large antimatter objects is convenient to separate 
into two different classes: clouds of gas and 
compact star-like or smaller but dense clumps of antimatter.
The boundary line between this two classes is determined by the
comparison of the mean free path of protons inside them, $l_p$,
and their size, $R_B$. If  $ l_p >R_B$ the annihilation of
antimatter in the cloud proceeds in all the volume of such B-bubble.
In the opposite case the annihilation takes place only on the surface.
The proton mean free path can be estimated as:
\be
l_p = \frac{1}{\sigma_{tot} n_{\bar p}} = 10^{24}\, cm \,
\left(\frac{cm^{-3}}{n_{\bar p}}\right)\, 
\left(\frac{barn}{\sigma_{tot}}\right)
\label{l-p2}
\ee
If the number density of antiprotons inside the bubble, $\bar n$, is
much larger (which is typically the case) than the number density
of protons in the background, i.e.
{${{ n_{\bar p} >> n_p}}$,} then it is possible that for
B-ball smaller than 
${{ l_{gal} = 3-10 \,\, kpc}}$ 
both limiting cases can be realized: 
volume annihilation ${ {l_{free}> R_B}}$, i.e. clouds, and
surface annihilation, $ l_{free} < R_B $, i.e. compact (stellar-like)
objects.

One should expect that typically an anti-cloud could not survive in a galaxy. 
It would disappear during
\be{
 \tau =  10^{15} \,\, sec\,\, 
\left(\frac{10^{-15}cm^3/s}{\sigma_{ann} v}\right)\,
\left(\frac{cm^{-3}}{n_p}\right),}
\label{tau}
\ee
if the supply of protons from the galactic gas is sufficient.
The proton flux into an anti-cloud is equal to:
\be{{
F = 4\pi l_c^2 n_p v = 10^{35}\,sec^{-1}\left(\frac{n_p}{cm^3}\right)
\left(\frac{l_c}{pc}\right)^2,
}\label{F}
}\ee
where $l_c$ is the cloud size, previously denoted as $R_B$. 
The total number of ${ \bar p}$ in the cloud is 
{${{ N_{\bar p} = 4\pi l_c^3 n_{\bar p}/3}}$.} The flux of protons
form the galactic gas is sufficient to destroy the anti-cloud in less 
than the universe age, i.e. $3\cdot {10^{17}}$ seconds, if:
\be{{
\left(\frac{n_{\bar p}}{cm^3}\right)
\left(\frac{l_c}{pc}\right) < 3\cdot 10^4
}}\label{n-bar-p}
\ee
Thus very large clouds might survive even in a galaxy. Almost surely 
they would survive in the halo.

In the case of volume annihilation, i.e. for ${ l_{free}^p> l_c}$ the 
number of annihilation per unit time and volume is
\be{
\dot n_p = v\sigma_{ann} n_p n_{\bar p}
\label{dot-n-p}
}\ee
The total number of annihilation per unit time is: 
{${ \dot N_p = 4\pi l_c^3\,\dot n_p /3}$.} The total number of 
${ \bar p}$ in the cloud is equal to:
{${{ N_{\bar p} = 4\pi l_c^3 n_{\bar p}/3}}$.} Comparing these two 
expressions we find the life-time (\ref{tau}) of the cloud.

The luminosity for {volume annihilation is equal to:
\be{{
L_\gamma^{(vol)} 
\approx
 10^{35}\,{\rm \frac{erg}{s}}\, 
 \left(\frac{R_B}{0.1\,{\rm pc}}\right)^3}} 
 {{\left( \frac{n_p}{{\rm 10^{-4}\,cm}^{-3}}\right)}}
{{ \left(\frac{n_{\bar p}}{10^4 {\rm cm}^{-3}}\right). 
}}\label{L-gamma-vol2}
\ee
and the flux of gamma rays on the Earth from anti-cloud at the distance
of d=10 kpc would be:
{${10^{-7}\gamma/{\rm s/cm}^2 }$} or {${10^{-5}{\rm Mev/\,s/cm}^2 }$ },
to be compared with cosmic background {${10^{-3}/{\rm MeV/s/cm}^2 }$.}
Still such annihilating cloud can be observed with a sufficiently
good angular resolution of the detector.

The compact stellar type objects for which ${{ l_s \gg l_{free}}}$
experience only the surface annihilation - all that hits the surface 
annihilate. There should be different sources of photons with 
quite different energies. The gamma-radiation from 
${{ \bar p p \rar pions}}$ and 
${{ \pi^0 \rar 2\gamma}}$ (${{ E_\pi \sim 300}}$ MeV) 
would have typical energies of hundreds MeV. The photons
from ${ e^+e^-}$-annihilation originating from ${{ \pi^\pm}}$-decays
${ \pi\rar \mu\nu}$, ${ \mu \rar e\nu\bar\nu}$, would be mostly below
100 MeV, while those from the "original" positrons in the B-ball would
create a pronounced 0.511 MeV line.

The total luminosity with respect to surface annihilation is 
proportional to the number density of protons in the Galaxy and to their
velocity, ${{ L_{tot} = 8\pi m_pl_s^2\,n_p v }}$. From this
we obtain:
\be{{{
L_{tot} 
\approx 10^{27}\,\frac{erg}{sec}\,
\left(\frac{n_p}{cm^3}\right)\left(\frac{l_s}{l_\odot}\right)^2, 
}\label{L-tot}
}}\ee
from which the fraction into gamma-rays is about 20-30\%.

\subsection{Antimatter from stellar wind}

Surprisingly the luminosity created by the annihilation of antiprotons
from the stellar wind may be larger than that from the surface annihilation.
The flux of particles emitted by an antistar per unit time can be written as: 
\be{{
\dot M = 10^{12} W\,g/sec
}}\label{dot-M}
\ee
where parameter $W$ describes the difference of matter emission
by solar type star and the anti-star under consideration:
${ W=\dot M /\dot M_{\odot}}$. For solar type antistar $W\approx 1$, while
for already evolved antistar $W\ll 1$.
If all ``windy'' particles (antiprotons and heavier antinuclei)
annihilate, the luminosity per antistar would be $ L= 10^{33} W $ erg/sec.

One sees that the luminosity of compact antimatter objects in the 
Galaxy is not large and it is not an easy task to discover them. However
such objects may have an anomalous chemical content which would be
an indication for possible antimatter. According to the discussed above
scenario of generation of cosmic antimatter objects they should have
anomalously large baryon-to-photon ratio. This leads to anomalous
abundances of light elements in this regions, for example such
domains should contain much less anti-deuterium and more anti-helium
than in the standard case with $\beta = 6\cdot 10^{-10}$. 
Moreover, some heavier primordial elements in the regions with high $\beta$
can be formed~\cite{anom-BBN}.
So the search for antimatter should start from a search of cosmic
clouds with anomalous chemistry. If such a cloud or compact object is found, 
one should search for a strong annihilation there. With 50\% probability
this may be, however, the normal matter with anomalous ${ n_B /n_\gamma}$ 
ratio, i.e. B-bubble with positive baryonic number.

Stellar wind and explosions of antistars would lead to enrichment
of the Galaxy with low energy antiprotons. The
life-time of ${ {\bar p}}$ with respect to annihilation in the 
Galaxy can be estimated as: 
\be{{
\tau = 3\cdot 10^{13}\,sec\,(barn/\sigma_{ann}\,v).
}}\label{tau-2}
\ee
The total number of antiparticles in a galaxy is determined by
the equation:
\be{{{
\dot {\bar N} = -\sigma_{ann}v\, n_p n_{\bar p} V_{gal} + S
}}\label{dot-bar-N}
}\ee
where ${ S}$ is the source, i.e.
{${{ S = W\epsilon (N_s/10^{12})\,10^{48}/sec}}  $,}
${ N_s}$ is the number of stars in the galaxy, $ \epsilon$ is the
fraction of antistars.
The stationary solution of the above equation is
\be{{
n_{\bar p} = \left(\frac{3\cdot 10^{-5}}{cm^{3}}\right)\epsilon W
\left(\frac{N_s}{10^{12}}\right)\,\left(\frac{barn}{ \sigma_{ann}v}\right).
}}\label{n-bar-p2}
\ee

{The number density of antinuclei is bounded by the density of
``unexplained'' ${ \bar p}$ and the fraction of antinuclei in
stellar wind with respect to antiprotons.} 
It may be the same as in the Sun but if antistars are old and evolved, 
this number may be much smaller.
Heavy antinuclei from anti-supernovae may be abundant but their
ratio to ${ \bar p}$ cannot exceed the same for normal SN.
{Explosion of anti-SN would create a large cloud of antimatter, which
should quickly annihilate producing vast energy - a spectacular event.}
{However, most probably such stars are already dead and SN might
explode only in very early galaxies or even before them.}

\subsection{Cosmic positrons}

Antistars can be powerful sources of low energy positrons.
The gravitational proton capture by an antistar is more 
efficient than capture of electrons because of a larger
mobility of protons in the interstellar medium. A positive charge
accumulated by the proton capture should be
neutralized by a forced positron ejection.
It would be most efficient in galactic center where ${n_p}$ is large.
The observed 0.511 MeV annihilation line must
be accompanied by wide spectrum $ {\sim 100}$ MeV radiation.

\subsection{Violent phenomena}

A collision of a star with an anti-star of comparable mass would lead
to a spectacular event of powerful gamma radiation similar to  
$\gamma$-bursters. The estimated energy release would be of the order of:
\be {{
\Delta E \sim 10^{48}\,erg\, \left(\frac{M}{M_\odot}\right)
\left(\frac{v}{10^{-3}}\right)^2
}}\label{Delta-E}
\ee
Since the annihilation pressure pushes the stars apart, the  
collision time would be quite short,${\sim 1}$ sec. The radiation
would be most probably emitted in a narrow disk but not in jets.

Another interesting phenomenon, though less energetic,  is a
collision of an anti-star with a red giant. In this case the compact 
anti-star would travel inside the red giant creating 
an additional energy source. It could lead to a change of color and 
luminosity. The expected energy release is
{${\Delta E_{tot} \sim 10^{38}}$ erg during
the characteristic time ${\Delta t \sim }$ month.} 

The transfer of material in a binary star-antistar 
system would lead to a very energetic burst of radiation
similar to a hypernova explosion. 

More difficult for observation and less spectacular effects include
the photon polarization. Since positrons are predominantly 
``right handed'', the same helicity is transferred to bremsstrahlung
photons. Indeed, neutron decay creates left-handed ${ e^-}$ and
antineutron creates right-handed positrons.
The first burst from SN explosion consists predominantly of
antineutrinos while that from anti-SN consists of neutrinos.

\subsection{Baryonic and antibaryonic dark matter}

The model considered above opens a possibility that all 
cosmological dark matter is made out of normal baryonic and
antibaryonic staff in the form of compact stellar-like
objects as early formed and now dead stars or primordial
black holes, either with mass near solar mass or much smaller,
e.g. near $10^{20}$ g.
 
Such objects could make all cold dark matter (CDM) in the universe
but in contrast to the usually considered CDM they are much heavier
and have a dispersed (log-normal) mass spectrum.
Very heavy ones with {${{ M>10^6 M_\odot}}$} which might 
exist on the high mass tail of the distribution
could be the seeds of large galaxy formation. Lighter stellar
type objects would populate galactic halos as usual CDM. 

The bounds on stellar mass object in the halo of the Galaxy
is presented in Fig. 3, taken from ref.~\cite{machos}.
\begin{figure}[t]
  \vspace{9.0cm}
  \includegraphics{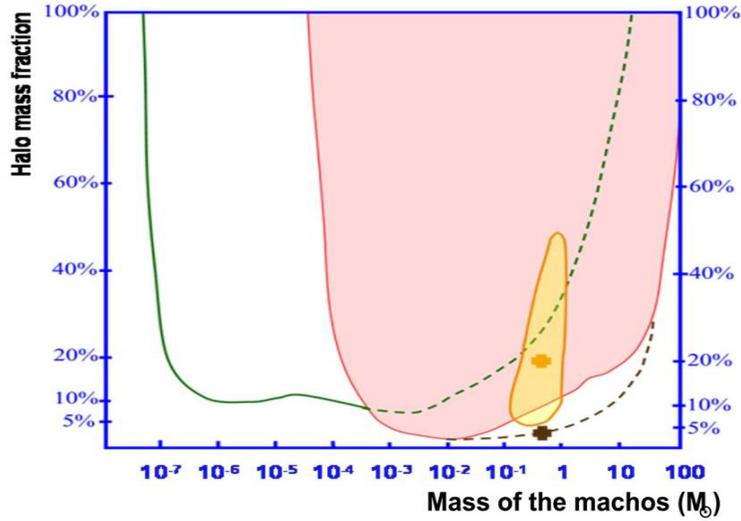}
  \caption{\it
    Micro-lensing bounds on compact objects in the
galactic halo as a function of their mass 
    \label{fig-machos} }
\end{figure}
No luminous stars are observed in the halo. It means that 
all high B compact objects are mostly already dead stars or PBH.
So the stellar wind must be absent. However, annihilation of background 
protons on the surface should exist and lead to gamma ray emission.

\subsection{Observational bounds}

The total galactic luminosity of the 100 MeV photons, $L_\gamma = 10^{39}$ erg/s,
and the flux of the $e^+e^-$--annihilation line, 
$F \sim 3\cdot 10^{-3}$ cm$^2$/s, allow to put the following
bound on the number of antistars in the Galaxy from the consideration of the
stellar wind:
\be
N_{\bar S} / N_S \leq 10^{-6} W^{-1}.
\label{N-bar-N-S}
\ee
It is natural to expect that $ {W\ll 1}$  because the primordial antistars 
should be already evolved.

From the bound on the antihelium-helium ratio (see e.g. review~\cite{picozza}) 
follows:
\be{{
N_{\bar S} / N_S  = (\bar{He}/He) \leq 10^{-6},
}}
\label{N-bar-N-S2}
\ee
if the antistars are similar to the usual stars, though they are
most probably not.

The only existing now signature in favor of cosmic antimatter is the 
observed 0.511 MeV photon line from galactic center and probably even 
from the galactic halo. However, other explanations are also possible
(for the list of references see~\cite{cb-ad-ap}).

\section{Conclusion}

{1. The Galaxy may possess a noticeable amount of antimatter.
Both theory and observations allow for that.}\\
2. Theoretical predictions are vague and strongly model dependent.\\
{3. Not only ${ ^4 \bar{He}}$ is worth to look for but 
also heavier anti-elements. Their abundances should be similar 
to those observed in SN explosions.} \\
{4. The regions with anomalous abundances of light elements suggest
that they consist of antimatter.}\\
{5. A search of cosmic antimatter has non-vanishing chance to be 
successful.}\\
6. Dark matter made of BH, anti-BH, and dead stars is a promising
candidate. There is a chance to understand why ${ \Omega_B =0.05}$ is 
similar by magnitude to ${ \Omega_{DM} = 0.25}$.

\end{document}